\documentclass[journal]{IEEEtran}
\usepackage{comment}
\usepackage{mathtools}
\usepackage{amsmath}
\usepackage{graphicx}
\usepackage{epstopdf}

\usepackage{hyperref} 
\usepackage{cite}
\usepackage{amsmath}
\usepackage{color}

\usepackage{graphicx}
\usepackage{mathptmx}      
\usepackage{latexsym}
\usepackage[ruled,vlined]{algorithm2e}

\usepackage[misc]{ifsym}
\usepackage{subfig}

\begin{document}

\title{Experience Dependent Formation of Global Coherent Representation of Environment by Grid Cells and Head Direction Cells}

\author{Taiping Zeng, XiaoLi Li, and Bailu Si
\thanks{T. Zeng is with Institute of Science and Technology for Brain-Inspired  Intelligence, Fudan University, Shanghai, China and Key Laboratory of Computational Neuroscience and Brain-Inspired Intelligence (Fudan University), Ministry of Education, China (e-mail:zengtaiping.ac@gmail.com).}
\thanks{X. Li, State Key Laboratory of Cognitive Neuroscience and Learning and IDG/McGovern Institute for Brain Research, Beijing Normal University, Beijing, China (e-mail:xiaoli@bnu.edu.cn).}
\thanks{B. Si is with School of Systems Science, Beijing Normal University, 100875, China (e-mail:bailusi@bnu.edu.cn).}
\thanks{Correspondence should be addressed to Bailu Si (bailusi@bnu.edu.cn).}
}


\maketitle

\begin{abstract}
The grid firing patterns are thought to provide an efficient intrinsic metric capable of supporting universal spatial metric for mammalian spatial navigation in all environments. 
However, whether spatial representations of grid cells in the entorhinal cortex are determined by local environment cues or form globally coherent patterns remains undetermined. 
To explore this underlying mechanism, here we proposed a possible theoretical explanation to describe connection between the neural space and the physical environment and transition from a local anchored to a global coherent representation according to relationship between grid phase distance between physical distance in the physical environment, and tested our method based on simultaneous localization and mapping (SLAM) system on a iRat rodent-sized robot platform in a rat-like maze. 
Our robotic exploration experiments show that the grid firing firstly is determined by local environment cues, and after self-correction with experience-dependence, the regular, continuous grid firing patterns tessellate the explored space. Head direction (HD) cells also show global patterns in our experimental results.
Our results support that grid firing patterns do provide a universal spatial metric for mammalian spatial navigation in complex environments.
The results in this study also provide insights into experience-dependent interactions between path integrative calculation of location in the entorhinal cortex and learned associations to the external sensory cues in the hippocampus, which is likely to be critical understanding spatial memory, even episodic memory.
\end{abstract}


%
\IEEEpeerreviewmaketitle

\section{Introduction}
\label{intro}

The cognitive map, a map-like internal representation of spatial environment, allows an animal to navigate effortlessly in the space~\cite{tolman_cognitive_1948}.
Grid cells in the medial entorhinal cortex (MEC) form a regular grid-like patterns spanning the explored environment~\cite{hafting_microstructure_2005}, whose periodicity allows to form highly effective spatial representations. These cells are thought to act as an intrinsic neural metric for mammals performing spatial navigation tasks in an environment~\cite{mcnaughton_path_2006,fiete_what_2008,buzsaki_memory_2013}. Grid cell activity is likely to provide a universal spatial metric in all environment~\cite{moser2014grid}.

Despite that, recent discoveries also show that due to various environmental features, grid firing patterns are distorted and fragmented. 
The grid cell firing patterns rescale in response to shrinking or stretching of the environment~\cite{barry_experience-dependent_2007}. 
Complex spatial environments are not represented holistically by the hippocampus and entorhinal cortex, but rather with discrete submaps that correspond to the environment's size and shape~\cite{derdikman_fragmentation_2009}. 
The hexagonal firing patterns of grid cells are rotated and deformed by geometric structures of the space~\cite{stensola_shearing-induced_2015}. 
Environmental geometry strongly and permanently affects the grid orientation, symmetry, scale, and homogeneity, and the symmetry of hexagonal grid firing patterns are permanently broken in a highly polarized environment, like trapezoid walls~\cite{krupic_grid_2015}.
Based on the above experimental results, grid firing patterns represent spatial environment in local discrete submaps way, not an universal spatial metric in all environments.

However, a regular, continuous grid firing pattern is required for an accurate and universal metric to uniformly cover the environment to be navigated.
Another recent exciting experiment shows that initially local sensory cues dominate grid firing patterns, and with self-correct experience, distortions and discontinuities were reduced to form the globally coherent grid firing patterns~\cite{carpenter_grid_2015}. Until now, computational models of grid cells mostly show the spatial firing rate map of integrating self-motion when mammals explores the spatial environment~\cite{mcnaughton_path_2006,fuhs_spin_2006, yoram_burak_accurate_2009,si_continuous_2014}. A coherent global representation is necessary to identify relative locations of all places in the environment. Still, this transition from a local to a global representation remains elusive~\cite{carpenter_grid_2015}.

In this paper, we proposed a method that is capable of transiting local firing patterns of grid cells and head direction cells to form globally coherent representation, which is required for a universal spatial metric in the complex spatial environment. For grid firing patterns, we first transformed coordinate systems of neural space to the same coordinate system with the physical environment, since the coordinate system of neural space is usually different from the physical environment. Then, the physical distance changes are calculated from the locations difference between positions before and after loop closures optimization. Grid cell firing rate adjustment in neural space is achieved by the physical distance changes based on periodic boundary conditions. 
Although global patterns of HD cells are rarely investigated in the neurobiological experiments, patterns of HD cells present similar features with global grid patterns in our study. We can also calculate global coherent HD firing patterns only from head direction phase changes between before and after sensory cues calibration.
We tested our method based on our previous SLAM system~\cite{zeng2017cognitive} on a iRat rodent-sized robot platform in a rat-like maze (iRat 2011 Australia dataset)~\cite{ball_irat:_2010,ball_openratslam:_2013}. Our experimental results show that grid and HD firing patterns anchor to local sensory cues before loop closures optimization, and after correction with experience, periodic grid patterns tile two-dimensional environment, and HD cells are only firing in one global direction in the all explored space. Our experimental results present a possible explanation to global coherent representations of grid cells and HD cells for navigation in the complex spatial environment~\cite{carpenter_grid_2015}.

In conclusion, the major contributions of this work are in the following. First, we proposed a method that transmits local environmental cues anchored representations to global coherent firing patterns both for grid cells and HD cells, which describes the connection between distance in the physical environment and grid phases distance in the neural space. Second, we implemented the method on the SLAM system, and global representations of grid cells and HD cells are presented in our results.
For the biological significances, we provided the first theoretical support that even in the complex spatial environment, coherent global patterns of grid cells and HD cells can be generated to act as an accurate and universal metric required by large-scale and long term spatial navigation~\cite{carpenter_grid_2015}, in spite of the geometry structure of the space~\cite{barry_experience-dependent_2007, derdikman_fragmentation_2009, krupic_grid_2015, stensola_shearing-induced_2015}.
Spatial view cells in the CA3 region of the hippocampus associated with environmental views~\cite{rolls_scientific_2017} may provide feedforward information to subiculum of the hippocampus for correcting HD cells firing patterns, then feedback information from CA3 and corrected HD pattern together achieve global coherent firing pattern of grid cells. Understanding this experience-dependent interactions between the hippocampus and the entorhinal cortex is likely to be critical in understanding spatial memory, even episodic memory~\cite{barry_experience-dependent_2007}. 

Our study is improved from the previous work as it provides grid and HD cells models, detailed implication of SLAM system~\cite{taiping_Neurobiol_2017}.
The rest of this paper proceeds as follows. Our methods are described in Section 2. Section 3 presents our results. Results and future works are discussed in Section 4. We give a brief conclusion in Section 5.

\section{Methods}
\label{Methods}
In this section, we first describe the Bayesian attractor neural network model on which global representations of grid cells and HD cells are based. Subsequently, the topological map is optimized as a non-linear least-squares problem. Then, we present the algorithm that transits grid cells and HD cells firing rate to form globally coherent representation. Finally, we introduce the implementation of our SLAM system.

\subsection{Bayesian Attractor Neural Network Model}
The Bayesian attractor neural network model is a model based on probabilistic methods to represent grid pattern and HD firing response~\cite{taiping_Neurobiol_2017}. We provide an overview of the HD and grid cells model, and the grid cell model is an extended HD cell model.
This model includes integrator cells and calibration cells. Integrator cells are presented to represent the head direction and position of the robot. Visual cues are modeled by calibration cells. Cue conflicts are modeled by mutual inhibition. Global inhibition leads to form a stable firing state. Bayesian integration eliminates cue uncertainties. 

\subsubsection{Head Direction Cell Model}
HD cell model represents the rotation of the robot, whose velocity input is the same as the angular velocity of the robot in the physical environment. The neural activity of the HD cell model is updated by attractor dynamics, vestibular cues integration, and visual cues calibration. The HD phase $\theta$ is in the ring manifold $[0,2\pi)$.

\paragraph{Attractor Dynamics}

The normal distribution is introduced to describe the integrator cell and calibration cell 
\begin{equation}
f(x) = \frac{1}{{\sigma \sqrt {2\pi } }}e^{{{ - \left( {x - \mu } \right)^2 } / {2\sigma ^2 }}}.
\label{eq:normalDistribution}
\end{equation}
The attractor dynamic feature is achieved by mutual inhibition and global inhibition between integrator cell and calibration cell. The continuous representation feature and stable bump with single peak existing over time are endowed by global inhibition and mutual inhibition. 

The global inhibition can be defined by 
\begin{equation}
\begin{split}
\displaystyle
\frac{1}{\sigma_{\text{cc}}^{t}{}^2} &= \frac{1}{\sigma_{\text{inte}}^{t-1}{}^2} + \frac{1}{\sigma_{\text{cali}}^{t-1}{}^2}\\
\frac{1}{\sigma_{\text{inte}}^{t}{}^2} &= E\,\,\frac{\frac{1}{\displaystyle \sigma_{\text{inte}}^{t-1}{}^2}}{\frac{1}{\displaystyle \sigma_{\text{cc}}^{t}{}^2} }\\
\frac{1}{\sigma_{\text{cali}}^{t}{}^2} &= E\,\,\frac{\frac{1}{\displaystyle \sigma_{\text{cali}}^{t-1}{}^2}}{\frac{1}{\displaystyle \sigma_{\text{cc}}^{t}{}^2} },
\end{split}
\end{equation}
where $\displaystyle \frac{1}{\sigma_{\text{inte}}^{t-1}{}^2}$ and $ \displaystyle \frac{1}{\sigma_{\text{cali}}^{t-1}{}^2}$ are the previous weight of integrator cell and calibration cell, respectively. $\displaystyle \frac{1}{\sigma_{\text{cc}}^{t}{}^2}$ is the sum of previous weight. $\displaystyle \frac{1}{\sigma_{\text{inte}}^{t}{}^2}$ and $ \displaystyle \frac{1}{\sigma_{\text{cali}}^{t}{}^2}$ are the current weight. $E$ is a constant, which is the total neural activity energy.

The mutual inhibition can be written as 
\begin{equation}
\begin{split}
\displaystyle
\frac{1}{\sigma_{\text{inte}}^{t}{}^2} &= \frac{1}{\sigma_{\text{inte}}^{t-1}{}^2} - 
\Delta_{\text{inte}}\,\,\frac{1}{\sigma_{\text{cali}}^{t-1}{}^2}\\
\frac{1}{\sigma_{\text{cali}}^{t}{}^2} &= \frac{1}{\sigma_{\text{cali}}^{t-1}{}^2} - \Delta_{\text{cali}}\,\,\frac{1}{\sigma_{\text{inte}}^{t-1}{}^2},
\end{split}
\end{equation}
where $\Delta_{\text{inte}}$ and $\Delta_{\text{cali}}$ are the mutual inhibition intensity to each other.

\paragraph{Vestibular Cues Integration}
Path integration is performed by shifting th mean of the normal distribution without bump deformation. Path integration can be implemented by 
\begin{equation}
\begin{split}
\mu_{\text{inte}}^{t} = \bmod (\mu_{\text{inte}}^{t-1} + \nu^{t} \Delta t, 2\pi)\\
\mu_{\text{cali}}^{t} = \bmod (\mu_{\text{cali}}^{t-1} + \nu^{t} \Delta t, 2\pi),
\end{split}
\end{equation}
where $\mu_{\text{inte}}^{t}$ and $\mu_{\text{cali}}^{t}$ are the mean of integrator cell and calibration cell, $\nu^{t}$ is current velocity, $\Delta t$ is time interval between $t$ and $t-1$. 

\paragraph{Visual Cues Calibration}
The neural activity of HD cells are calibrated by familiar sensory cues. When a new view is received from camera, this new view as a view template associated with a new local view cell corresponds to current HD pattern by a strong link. When a familiar view comes, the local view cell can be reactivated. Energy can be injected into HD cells network through learned link. The energy injection can be written as 
\begin{equation}
\begin{split}
\displaystyle
\frac{1}{\sigma_{\text{cali}}^{t}{}^2} &= \frac{1}{\sigma_{\text{cali}}^{t-1}{}^2} + \frac{1}{\sigma_{\text{inject}}^{t}{}^2}\\
\mu_{\text{cali}}^{t} &= \bmod \left(\left( \frac{\mu_{\text{cali}}^{t-1}}{\sigma_{\text{cali}}^{t-1}{}^2} + \frac{\mu_{\text{inject}}^{t}}{\sigma_{\text{inject}}^{t}{}^2} \right) \,\, \sigma_{\text{cali}}^{t}{}^2, 2\pi\right),
\end{split}
\end{equation}
where $\displaystyle \frac{1}{\sigma_{\text{inject}}^{t}{}^2}$ is the intensity of the current injected energy, $\displaystyle \mu_{\text{inject}}^{t}$ is the injected location on the one dimensional neural manifold of HD cells.

\paragraph{Phase Estimation}
The current HD phase can be calculated from the integrator cell and the calibration cell, whose probabilistic distribution can be described by
\begin{equation}
\begin{split}
\displaystyle
\frac{1}{\sigma_{\text{cc}}^{t}{}^2} &= \frac{1}{\sigma_{\text{inte}}^{t}{}^2} + \frac{1}{\sigma_{\text{cali}}^{t}{}^2}\\
\mu_{\text{cc}}^{t} &= \bmod \left(\left( \frac{\mu_{\text{inte}}^{t}{}}{\sigma_{\text{inte}}^{t}{}^2} + \frac{\mu_{\text{cali}}}{\sigma_{\text{cali}}^{t}{}^2} \right) \,\, \sigma_{\text{cc}}^{t}{}^2, 2\pi\right),
\end{split}
\end{equation}
where $\displaystyle \frac{1}{\sigma_{\text{cc}}^{t}{}^2}$ and $\displaystyle \mu_{\text{cc}}^{t}$ are the estimated wight and HD phase, respectively. If $\displaystyle |\mu_{\text{cc}}^{t} - \mu_{\text{cali}}^{t}| < \text{Threshold}$, the decision that the robot enters a familiar environment is made, and HD phase $\displaystyle \mu_{\text{cc}}^{t}$ is assigned to $\displaystyle \mu_{\text{inte}}^{t}$. If it not meets the condition, it would go to the next cycle.

\subsubsection{Grid Cell Model}
We expand the ring manifold of the HD cell model to the torus manifold of the grid cell model. We adopt the same mechanism of HD neural system for location calculation. 
The integrator cell and calibration cell in the torus manifold can be described as
\begin{equation}
\displaystyle
f(x,y) = \frac{1}{{2\pi \sigma_x \sigma_y}}e^{ - [ \left( {x - \mu_x } \right)^2 / {2\sigma_x ^2 } + \left( {y - \mu_y } \right)^2 / {2\sigma_y ^2 } ] }.\label{eq:2DnormalDistribution}
\end{equation}
Same mechanisms in HD cell model can be used to grid cell representation integrating linear velocity and sensory cues, which also include attractor dynamics, vestibular cues integration, visual cues calibration, phase estimation.
We will not go further into the issue here.

\subsection{Graph-Based Non-Linear Least Squares Optimization for Topological Map}
A constrained robust non-linear least squares approach is employed to optimize topological map  in a graph~\cite{zeng_compactmapping_2017}. The pose of the robot is modeled as a node. The spatial constraint between nodes is modeled as a link. New constraints can be easily considered by adding new residuals. Then, Ceres Solver~\cite{agarwal2012ceres} can be adopted to compute an optimized solution to
\begin{equation}
\begin{split}
\displaystyle
\min_{\mathbf{e}} &\quad \frac{1}{2}\sum_{i,j} \rho_i\left(\left\|f_i\left(e_{i},e_{j},e_{ij}\right)\right\|^2\right),
\end{split}
\end{equation}
where, $\mathbf{e}$ including all nodes $e_i$ and $e_j$ is optimized given links $e_{ij}$, $e_i = (x_i,y_i,\theta_i)$ and $e_j = (x_j,y_j,\theta_j)$ are the node state. $e_{ij} = (x_{ij},y_{ij},\theta_{ij})$ describes the constraint from $e_i$ to $e_j$.  $\rho_i\left(\left\|f_i\left(e_{i},e_{j},e_{ij}\right)\right\|^2\right)$ is a residual block, where $f_i(\cdot)$ is a cost function. $\rho_i$ is a loss function, i.e. Huber Loss. Loss function can decrease the influence of outliers during the process of the global optimization for topological map. Cost function $f_i(\cdot)$ for a pair of nodes can be further described by 
\begin{gather}
\begin{split}
f_i\left(e_{i},e_{j},e_{ij}\right) 
&= \begin{bmatrix} 
e_j - e_i - e_{ij} 
\end{bmatrix} 
= \begin{bmatrix} 
x_j - x_i - x_{ij} \\ 
y_j - y_i - y_{ij} \\ 
\theta_j - \theta_i - \theta_{ij} 
\end{bmatrix}\\
&= \begin{bmatrix}
x_j - x_i - d_{ij} \cdot \cos(\theta_i + \text{heading\_rad}) \\ 
y_j - y_i - d_{ij} \cdot \sin(\theta_i + \text{heading\_rad}) \\ 
\theta_j - \theta_i - \text{facing\_rad}
\end{bmatrix}, \\
\text{s.t.} &\quad -\pi \le \theta_i < \pi ,\\
&\quad -\pi \le \theta_j < \pi,
\end{split}
\end{gather}
where $e_{ij}$ describes distance between $e_i$ and $e_j$, heading radians $\textit{heading\_rad}$, and facing radians $\textit{facing\_rad}$, $d_{ij}$ is the distance between $e_i$ and $e_j$. Due to the existence of relative angle radians when visual template matching, heading radians and facing radians are not the same value~\cite{ball_openratslam:_2013}. Values of $\theta_i$ and $\theta_j$ are limited to a certain range, which belongs to $[-\pi,\pi)$.

\subsection{Head Direction and Grid Firing Patterns Correction with Experience}
Mammals can explore long distances for forage, and then unmistakably return to their home. The familiar views in their home can calibrate their cognitive map in their brains to form a global coherent map to ensure that they do not get lost.

In our SLAM system, local view cells are used to represent a distinct visual view in the environment. A novel view leads to create a new local view cell and associate this current view to this new local view cell. Plus, that local view cell is excitingly linked to the current grid and HD cell neural activity patterns. Then, a new experience is created, defined by current grid cells, HD cells, and local view cells neural activity states.
When the robot sees the familiar view again, the activated local view cell injects energy into grid cells and HD cells neural networks through excitatory link. Given that each view is associated with a different discrete local view cell, for the long sequences of familiar views, the familiar views are successively recalled in the correct order~\cite{rolls_scientific_2017}, which would lead to gradually restore grid and HD cells neural activity states over time in the neural space. Even so, since odometry error accumulates and restoration of grid and HD patterns occurs, the phases of grid cells and HD cells in the neural space becomes discontinuous. Every close two adjacent local view cells actually associate two physical positions of the robot at great distances. Non-linear least squares is introduced to optimize experience map~\cite{agarwal2012ceres}. 

Now, although we optimize experience locations in the physical environment and restore current grid and HD patterns, all past experienced neural activities of grid and HD cells remain the same. 
Next, experience-dependent adjustment of HD cell and grid cell neural activity states is introduced to form global HD and grid patterns, according to all physical experience position changes between before and after graph optimization of cognitive map.

\subsubsection{Head Direction Cells}
\label{method_HDC}
In the ring attractor network of HD cells, as head direction in both neural space and physical environment remains the same scale, the adjusted HD phase can be directly calculated from head direction changes between before and after optimization in the physical environment. The adjusted HD phase can be written as  
\begin{equation}
\displaystyle
\theta_{n} = \theta^{\prime}_{n} + \bmod (\theta_{p\_opt} - \theta_{p\_org}, 2\pi),
\end{equation}
where $\theta_{p\_opt}$ and $ \theta_{p\_org}$ are head directions in the physical environment before and after optimization, respectively. $\theta^{\prime}_{n}$ is the HD phase before adjustment. $\theta_{n}$ is the HD phase after adjustment.

\subsubsection{Grid Cells}
In the torus attractor network of grid cells, the wrapping of network edges can map an infinite physical environment in theory. Usually, neural space of grid cell network has different coordinate system with the reference physical environment. The spatial wavelength of grid fields expressed by grid cells is determined by the gain of the velocity input~\cite{mcnaughton_path_2006}. Plus, considering the periodicity of torus manifold, the position changes between before and after optimization in the physical environment can be used to calculate phase changes of grid cells network in the neural space. The adjusted grid phase in the two-dimensional environment is given by 
\begin{equation}
\displaystyle
\begin{split}
\begin{bmatrix} 
x_{n} \\
y_{n} \\
\end{bmatrix}
=
&\begin{bmatrix} 
\cos(\theta) & -\sin(\theta)\\ 
\sin(\theta) & \quad\cos(\theta)\\ 
\end{bmatrix}
\begin{bmatrix} 
x^{\prime}_{n} \\
y^{\prime}_{n} \\
\end{bmatrix}
\\
&+\bmod
(
\begin{bmatrix} 
x_{p\_opt} - x_{p\_org} \\ 
y_{p\_opt} - y_{p\_org} \\ 
\end{bmatrix} \cdot v_{scale}
, 2\pi
),
\end{split}
\end{equation}
where $(x_{p\_org}, y_{p\_org})$ and $(x_{p\_opt}, y_{p\_opt})$ are positions of the  robot before and after optimization, respectively. $v_{scale}$ is the gain of velocity input. $\theta$ is angle of rotation transformation from reference coordinate system of physical environment to coordinate system of neural space.
$(x^{\prime}_{n}, y^{\prime}_{n})$ is the phase of grid cells network before adjustment. $(x_{n}, y_{n})$ is the adjusted phase of grid cells.
This equation also shows the connection between physical distance and grid phase distance.

\subsection{Implementation of SLAM System}

\begin{figure*}[!ht]
\centering
\includegraphics[width=3.5in]{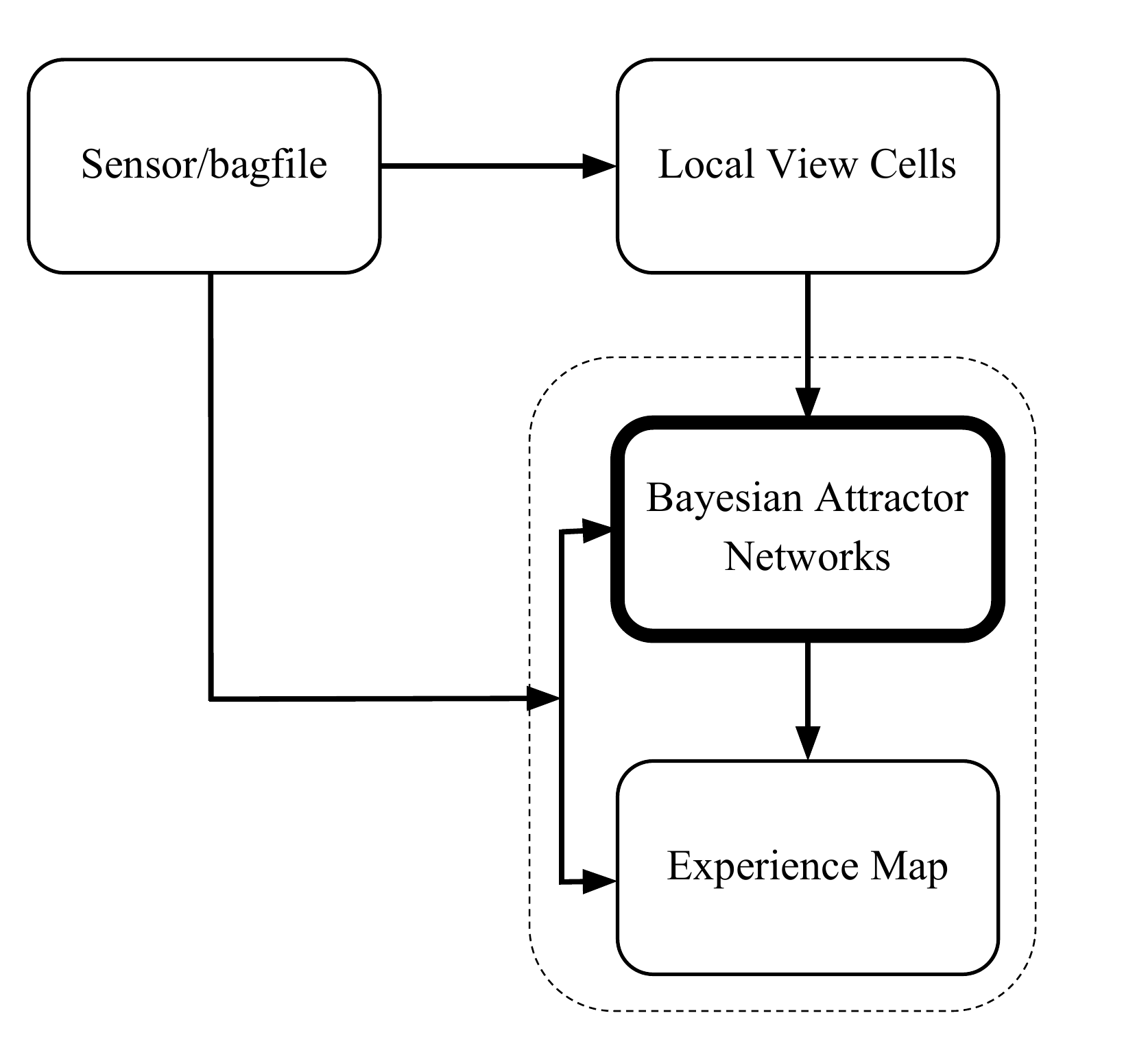}
\caption{The software architecture of the SLAM system. Images and odometry information are provided by the sensor/bagfile node. Whether the current view is familiar or not is determined by the local view cells node. The grid cells and HD cells model is implemented in the Bayesian attractor network node, which performs path integration and make decisions of loop closures. The experience map node achieves the global graph optimization of the experience map. }
\label{fig_node_structure}
\end{figure*}

Our SLAM system is run in Robot Operating System (ROS) Indigo on Ubuntu 14.04 LTS (Trusty), which is implemented in C++ language. The software architecture of the SLAM system is organized into four nodes shown in Fig.~\ref{fig_node_structure}. We reuse the local view match algorithm in openRatSLAM system~\cite{ball_openratslam:_2013}.

The sensor/bagfile node provides images and odometry information as inputs for our SLAM system.
The local view cell node compares the current image with view templates to determine whether the current view is familiar or not. Then, the calibration current is provided to the Bayesian attractor network node. 
The Bayesian attractor network node integrates the movement and sensory information by modeling grid cells and HD cells neural responses. This node also makes decisions about experience map node and link creation.
The experience map node builds the topological map and optimizes this map by graph-based non-linear least-squares approaches.

\section{Experimental Results}
\label{results}

\begin{figure*}[!htbp]
\centering
\includegraphics[width=0.8\textwidth]{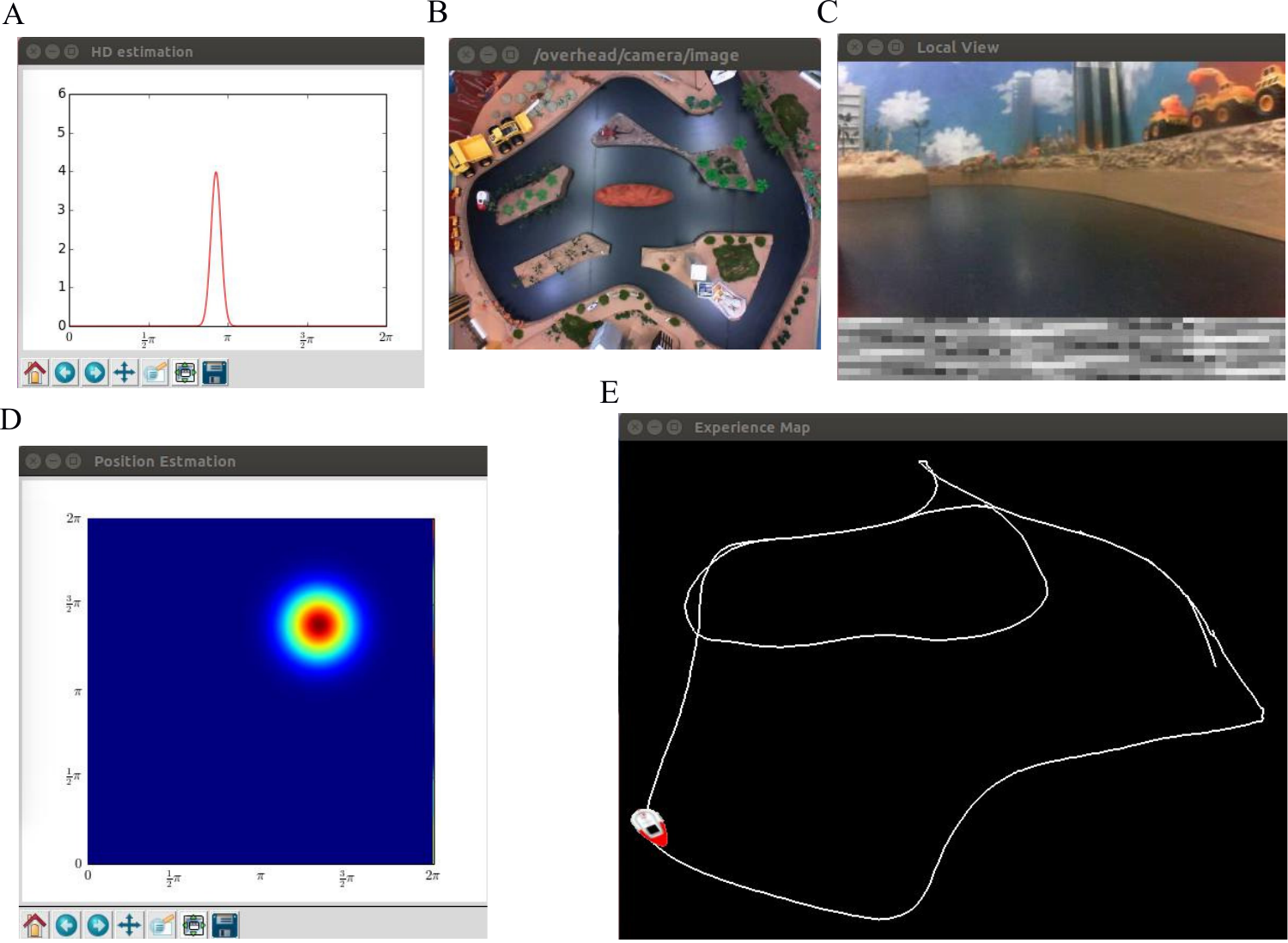}
\caption{Screenshots of SLAM system for iRat Australia dataset. (A) Neural activities of HD cells; (B) overhead image; (C) Neural activities of grid cells; (D) Input visual scene(top); the local view template and the matched template (bottom); (E) Experience map.}
\label{fig_iratrunsystem}
\end{figure*}

In this section, we tested our method on a publicly available open-source dataset, iRat Australia dataset. A small mobile robot, called Intelligent Rat animat technology (iRat), is a tool to investigate spatial navigation and cognition for robotics and neuroscience teams for interdisciplinary studies. iRat is similar to a rodent in size and shape. Camera images obtained by web camera, odometry messages, and overhead images are provided by the iRat ROS bag. Screenshots of our running SLAM system for iRat Australia dataset is shown in Fig.~\ref{fig_iratrunsystem}.

We mainly want to present global coherent patterns of HD and grid cells compared with local anchored patterns determined by local sensory cues in the experimental results. Some detailed results are also presented to further describe global patterns. No closed-loop firing rate maps show that without loop closures, though pure path integration can form global patterns, it can not generate a correct cognitive map. Different HD and grid cells firing rate maps are selected to show HD phases and grid phases in the results. Different gain of velocity inputs lead to global grid patterns with different grid spacings. The evolution of grid firing maps is described for the interval of one-quarter of the dataset. 

\subsection{Local Anchored and Global Coherent Patterns of Head Direction and Grid cells}
\begin{figure*}[!htb]
\centering
\includegraphics[width=0.6\textwidth]{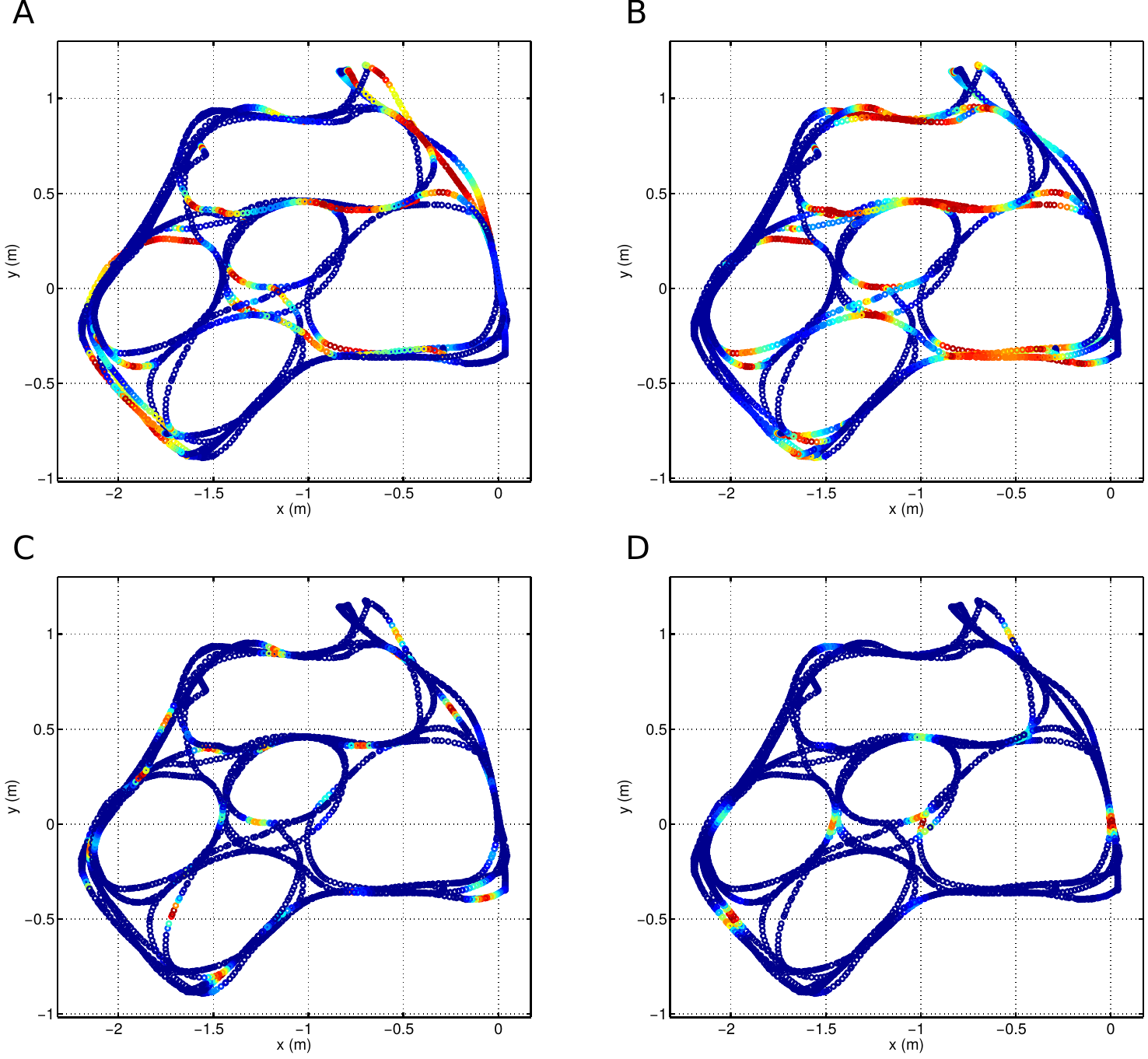}
\caption{Local anchored and global firing rate maps of HD and grid cells. In each panel, the firing rate is plotted at the locations in the experience map. The same jet colormap is used to color-code firing rate. The color scale is from blue (silent) to red (peak rate).
(A) The firing rate map of HD cells at both 0 and $\pi$ in the ring manifold, presents local anchored patterns. (B) The global HD patterns at both 0 and $\pi$ in the ring manifold after firing rate adjustment. (C) The local anchored grid pattern at $(0,0)$ in the torus manifold. (D) The global coherent grid pattern at $(0,0)$ in the torus manifold.}
\label{fig_localglobalpatterns}
\end{figure*}

To compare the local anchored firing patterns of HD and grid cells with global coherent patterns, firing rate maps are presented in Fig.~\ref{fig_localglobalpatterns}. Fig.~\ref{fig_localglobalpatterns}A and Fig.~\ref{fig_localglobalpatterns}C show the local anchored patterns of HD and grid cells, respectively. Although all HD firing patterns are not consistent in Fig.~\ref{fig_localglobalpatterns}A, the single HD firing field in one direction is always represented by the same firing rates. This HD firing pattern anchors to local sensory cues, formulated by self-motion during path integration. In Fig.~\ref{fig_localglobalpatterns}B, firing rate of HD cells are adjusted by methods in section~\ref{method_HDC}. The global HD patterns are only firing in the horizontal direction. 

For grid patterns, here, we map the physical environment onto an untwisted torus manifold, which would form square periodic firing fields in the experience map. Fig.~\ref{fig_localglobalpatterns}C shows the local anchored grid patterns. Since determined by local sensory environment cues, grid firing fields are scattered in the experience map. However, grid firing rate always increases first and then decreases gradually. As the periodic boundary conditions of the grid pattern in the torus manifold, the grid cell fires at multiple distinct locations in the all experience map. After firing rate adjustment, the regular, continuous grid pattern is presented in Fig.~\ref{fig_localglobalpatterns}D, which reflects absolute positions in the spatial space. Fig.~\ref{fig_localglobalpatterns}D shows that the grid pattern is at (0,0) in the torus manifold, and grid spacing is $0.5 m$, which is scaled by the gain of velocity input.

\subsection{No closed-loop and closed-loop Head Direction and Grid Cell Patterns}
\begin{figure*}[!htb]
\centering
\includegraphics[width=0.6\textwidth]{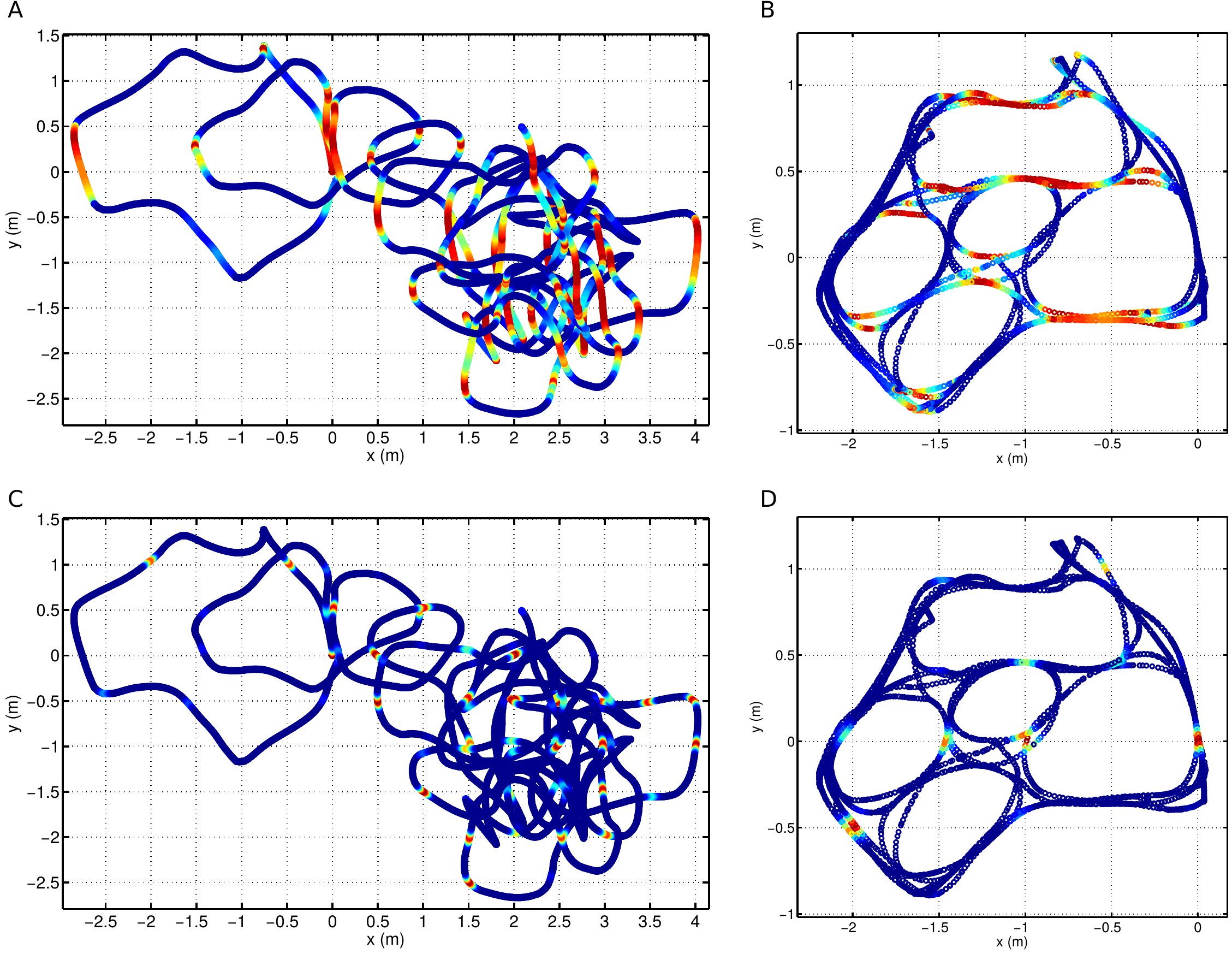}
\caption{No closed-loop and closed-loop HD and Grid Cell Patterns. The trajectories of (A) and (C) are formed by path integration without loop closures. (A) The firing rate map of HD cells at both $\frac{\pi}{2}$ and $\frac{3\pi}{2}$ in the ring manifold. (B) After loop closures and firing rate adjustment, the global HD firing patterns show in the corrected experience map. (C) The square grid patterns show in the firing rate map generated by path integration without loop closures. (D) The global grid patterns.}
\label{fig_noloopclosedpatterns}
\end{figure*}

To show that global patterns of grid cells and HD cells can be produced only by path integration, if there is no any odometry accumulated error, firing rate maps with no closed-loop is presented in Fig.~\ref{fig_noloopclosedpatterns}. In Fig.~\ref{fig_noloopclosedpatterns}A, HD cells fire at only two opposite directions, which is globally coherent. The firing rate map in Fig.~\ref{fig_noloopclosedpatterns}A has similar HD patterns in Fig.~\ref{fig_noloopclosedpatterns}B. In Fig.~\ref{fig_noloopclosedpatterns}C, global grid patterns are shown in the experience map also without loop closures and firing rate adjustment. Fig.~\ref{fig_noloopclosedpatterns}C and Fig.~\ref{fig_noloopclosedpatterns}D have the same grid spacing $0.5m$.
This can also show that without firing rate adjustment, global firing patterns with correct experience map, like Fig.~\ref{fig_noloopclosedpatterns}C and D, can not be achieved. 

\subsection{Global Head Direction and Grid Cell Patterns with Different Phases}
\begin{figure*}[!htbp]
\centering
\includegraphics[width=0.5\textwidth]{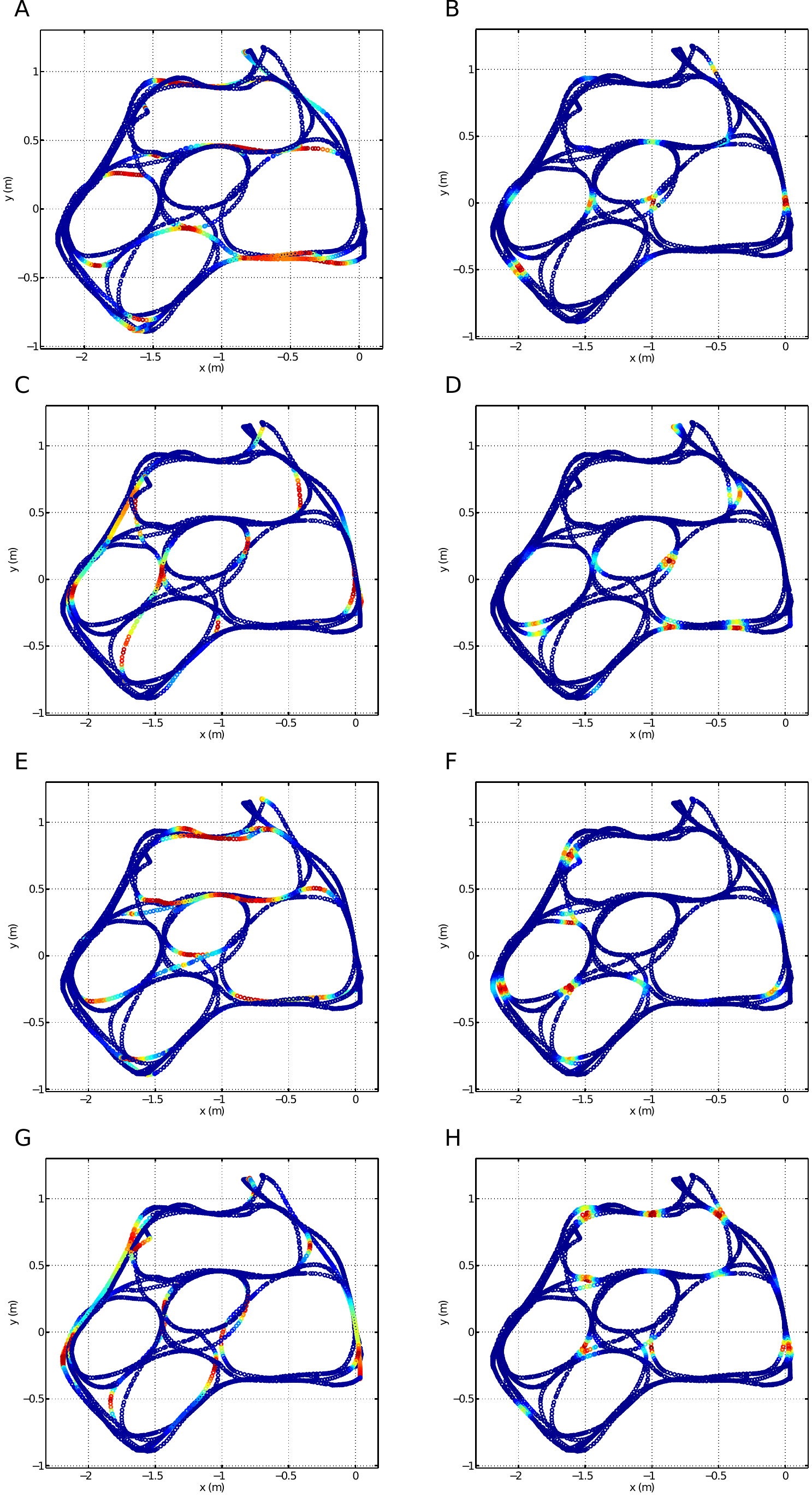}
\caption{Global HD and grid patterns with different phases. (A), (C), (E), and (G) show firing rate maps of HD cells at $0$, $\frac{\pi}{2}$, $\pi$, and $\frac{3\pi}{2}$ in the ring manifold, respectively. (B), (D), (F), and (H) show firing rate maps of grid cells at $(0,0)$, $(\frac{\pi}{2},\frac{\pi}{2})$, $(\frac{3\pi}{2},\pi)$, and $(2\pi,\frac{3\pi}{2})$, respectively.}
\label{fig_differentcellpatterns}
\end{figure*}

To show more detailed information about HD and grid cell patterns, the firing rate map of different HD cells in the ring manifold and grid cells in the torus manifold are displayed in Fig.~\ref{fig_differentcellpatterns}.
HD cells at $0$, $\frac{\pi}{2}$, $\pi$, and $\frac{3\pi}{2}$ in the ring manifold, represent four different directions shown in Fig.~\ref{fig_differentcellpatterns}A, C, E, and G, respectively. We also select four different grid cells in the torus manifold, i.e. $(0,0)$, $(\frac{\pi}{2},\frac{\pi}{2})$, $(\frac{3\pi}{2},\pi)$, and $(2\pi,\frac{3\pi}{2})$, to show different grid phases in Fig.~\ref{fig_differentcellpatterns}B, D, F, and H, respectively. 

\subsection{Global Grid Cell Patterns with Different Grid Spacings}
\begin{figure*}[!htbp]
\centering
\includegraphics[width=0.35\textwidth]{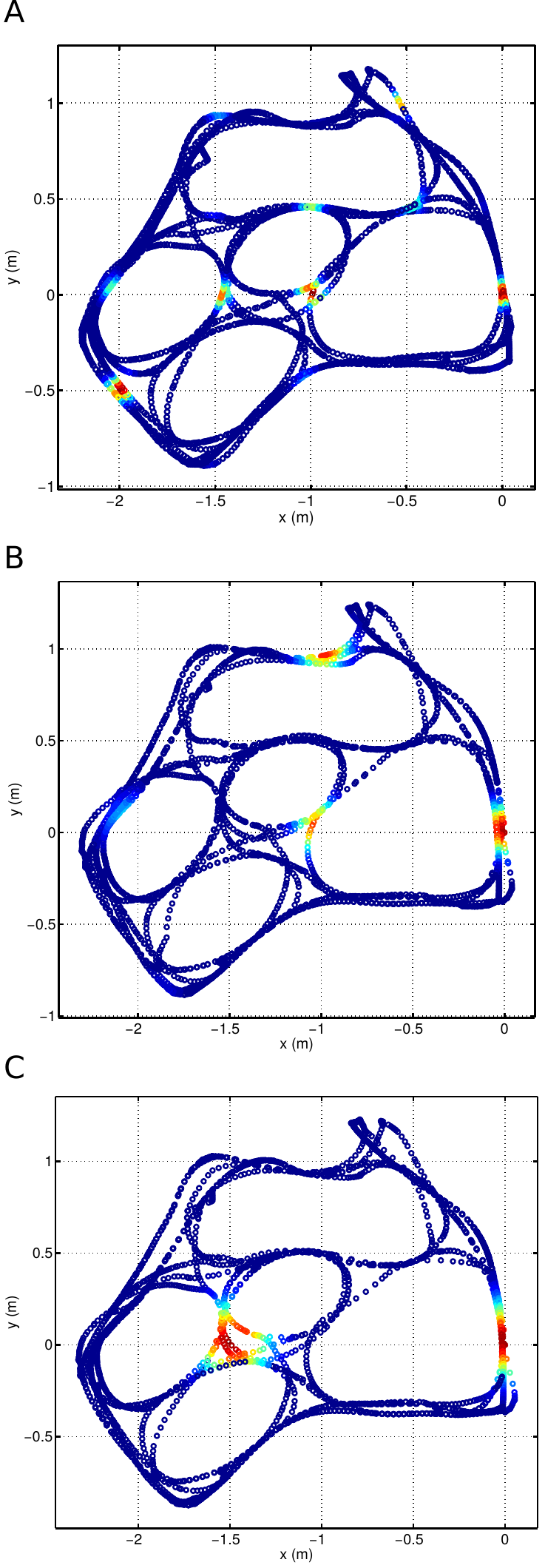}
\caption{Global grid patterns with different grid spacings. (A), (B), and (C) show firing rate maps with grid spacing $0.5m$, $1.0m$, and $1.5m$, respectively.}
\label{fig_differentgridscales}
\end{figure*}

Since grid cells have multiple dimensions of variation, we already show the different grid phases in Fig.~\ref{fig_differentcellpatterns}. We, here, present grid patterns with different grid spacings in Fig.~\ref{fig_differentgridscales}. The firing rate maps with grid spacing $0.5m$, $1.0m$, and $1.5m$ shown in Fig.~\ref{fig_differentgridscales}A, B, and C, respectively. All grid patterns start at position $(0,0)$ in the experience map. The grid spacings are determined by the gain of velocity inputs.

\subsection{Grid Cell Patterns of Firing Rate Map Evolution}
\begin{figure*}[!htbp]
\centering
\includegraphics[width=0.7\textwidth]{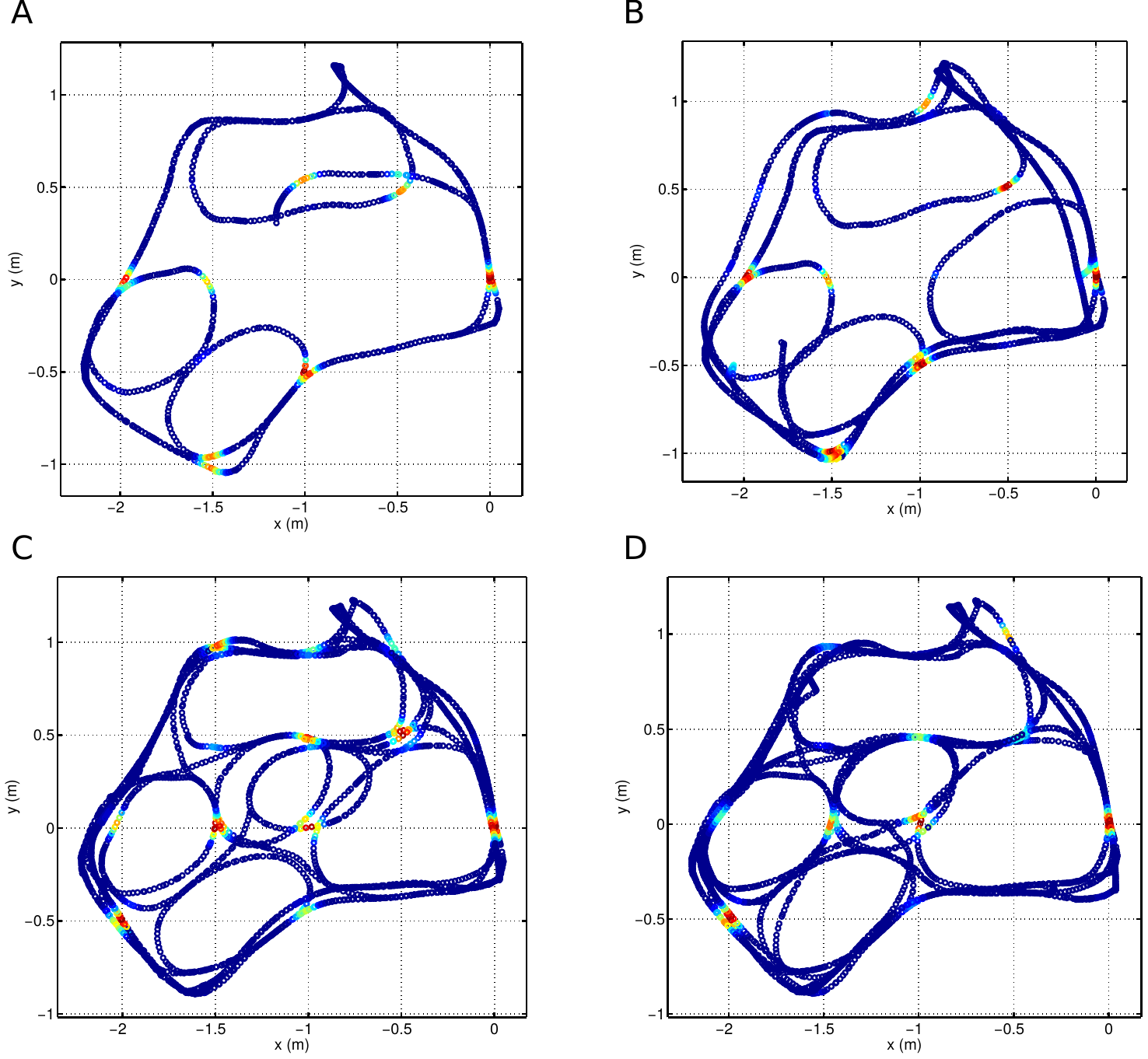}
\caption{Firing rate map evolution for grid cell patterns. (A), (B), (C) and (D) show the firing rate map of grid cells for each interval of one quarter of the dataset. Although large localization errors occur regularly, firing rate adjustment can alway maintain global coherent grid firing patterns.}
\label{fig_gridpatternevolution}
\end{figure*}

Fig.~\ref{fig_gridpatternevolution} presents the grid patterns in the evolution of the experience map over time.
For each interval of one-quarter of the dataset, we show the corresponding firing rate map of grid cells in Fig.~\ref{fig_gridpatternevolution}A, B, C, and D. 
Due to the poor odometry, large localization errors occur regularly, which are corrected by loop closures. Even if odometry errors accumulate, the global coherent grid firing patterns can also be formed by appropriate firing rate adjustment. The final experience map in Fig.~\ref{fig_gridpatternevolution}D is similar to the ground truth in Fig.~\ref{fig_iratrunsystem}B.

\section{Discussion}
\label{discussion}
We proposed an algorithm that can form a coherent global pattern acting as an accurate and universal spatial metric. We demonstrated that this algorithm can transit grid and HD cell patterns from a local anchored to a global coherent representation, which provides the first theoretical evidence for grid firing patterns capable of supporting universal metric spatial navigation in a complex environment~\cite{carpenter_grid_2015}. We tested this algorithm in the rat-like maze (iRat 2011 Australia dataset)~\cite{ball_openratslam:_2013} based on our SLAM system.

Vestibular calculations for mammals would be prone to significant accumulated errors. Grid cells can not form a regular and continuous, but distorted and discontinuous patterns, due to geometry features of space~\cite{barry_experience-dependent_2007, derdikman_fragmentation_2009, krupic_grid_2015, stensola_shearing-induced_2015}.
To be useful for mammals spatial navigation, grid cells cannot only respond to self-motion cues from vestibular, like current grid cells models~\cite{mcnaughton_path_2006,fuhs_spin_2006, yoram_burak_accurate_2009,si_continuous_2014}. Grid cells must anchor to external reference frames~\cite{edvard_moser_nobel_lecture2014}. Our main contribution is that our proposed transition method with our Bayesian attractor network can adjust grid cells and HD cells firing rate to generate the coherent global representation shown in Fig.~\ref{fig_localglobalpatterns}, which is required for them to act as an effective spatial metric. The adjusted HD and grid firing activity can consistently anchor to external reference frames. This global patterns provide necessary information for mammals to identify the relative positions of all locations in the environment.
This firing rate adjustment is actually an experience-dependent interaction between sensory and path integration information, which provides an accurate self-localization may be mediated by the entorhinal grid cells. Self-correction with experience-dependent interaction leads to form the coherent global grid and HD patterns spanning all the environment. 
For the further implication, this experience-dependent interaction is one of interaction between the hippocampus and entorhinal cortex, which is very likely to provide an understanding of spatial memory~\cite{barry_experience-dependent_2007}.

In Fig.~\ref{fig_localglobalpatterns} and Fig.~\ref{fig_noloopclosedpatterns}, three different firing map states, only path integration, local anchored representation, global representation, are presented to show that if path integration is completely accurate, global pattern can be directly presented shown in Fig.~\ref{fig_noloopclosedpatterns}; if experience map in the physical environment are optimized to achieve a corrected map, the representation of grid and HD cells in the neural space are also need to adjust their firing rate states. To detailedly describe global grid patterns, grid phase, and grid spacing features are shown in Fig.~\ref{fig_differentcellpatterns} and Fig.~\ref{fig_differentgridscales}, respectively. Fig.~\ref{fig_gridpatternevolution} shows that global patterns alway exist in all mapping process. As the robot only explores the path not the whole two-dimensional plane of environment, global pattern mapping onto torus manifold is far more easy to observe than onto twisted one.

Further, grid cells appeared to shift their firing rate states gradually and continuously, rather than undergo abrupt changes in representations~\cite{carpenter_grid_2015}. However, as the robot travels environment, this continuous transformation probably would not happen in our experiments. The probable reason is that although mammals need to explore the environment many times to get familiar with surroundings, the robot can remember all experience features at one time. 
Plus, considering that the interaction between sensory information and path integration is likely to be a probabilistic method~\cite{knight_weighted_2012,page_theoretical_2014,zhang_decentralized_2016}, Bayesian attractor network is used in our experiments.

Several limitations still remain to be studied. First, place cells are not included in our models. Second, the transition from a local to global representation is not continuous.

In the future, we plan to further investigate the interaction between the hippocampus and entorhinal cortex to understand spatial memory. Place cells integrating multiple grid cell modules would be used to study experience-dependent interaction through back projections via CA1, subiculum to the entorhinal cortex. Plus, the underlying mechanism of global grid cell representation can be probably formed by attractor neural network to shift the hexagonal firing fields of the grid cells gradually and continuously.

\section{Conclusion}
\label{conclusion}
In a word, a method is proposed to transit grid cells and HD cells from a local anchored to a global coherent representation by experience-dependent correction in the complex environment, which is required for mammals to act as an accurate and universal metric capable of supporting universal metric spatial navigation in all environment.
This transition mechanism reflects an interaction between path integrative calculation of location in the entorhinal cortex and learned associations to the external sensory cues in the hippocampus, which probably facilitates understanding spatial cognition and memory, more generally, human episodic memory.


\ifCLASSOPTIONcaptionsoff
  \newpage
\fi

\bibliographystyle{IEEEtran}
\bibliography{IEEEabrv,globalpattern}





\vfill



\end{document}